\documentclass[english]{article}
\usepackage[T1]{fontenc}
\usepackage[latin9]{inputenc}
\usepackage{color}
\usepackage{babel}
\usepackage{textcomp}
\usepackage{amsmath}
\usepackage{amssymb}
\usepackage[unicode=true,pdfusetitle,
 bookmarks=true,bookmarksnumbered=true,bookmarksopen=true,bookmarksopenlevel=1,
 breaklinks=false,pdfborder={0 0 1},backref=false,colorlinks=true]
 {hyperref}
\usepackage{breakurl}

\makeatletter
\@ifundefined{date}{}{\date{}}
\makeatother

\begin{document}

\title{Canonical quantization in a spinor substructure of Minkowski space}

\author{Kaare Borchsenius}
\maketitle
\begin{abstract}
We factorize the space-time coordinates of Minkowski space into Weyl
spinors with components in a split Clifford algebra. Poisson brackets
are defined for spinor-valued canonical variables and applied to the
quantization of point particles and strings. In particular, we obtain
the Lorentz algebra for the quantum string, and show that the string
supports both integral and half-integral spin states. The Clifford
algebra is augmented with the octonions through an R-algebra tensor
product, and we apply the results of Manogue, Schray and Dray on octonionic
Lorentz transformations to obtain a Lorentz invariant string action
in ten dimensions.
\end{abstract}

\section{Introduction and mathematical preliminaries}

At first glance there seems to be no connection between the quantum
interference between alternative space-time paths and the double homomorphism
$SL(2,\mathbb{C})\rightrightarrows SO(1,3)$ of the Lorentz group.
However, if space-time is equipped with a spinorial substructure,
the interference terms in the transition probabilities can be interpreted
as single amplitudes in the underlying $SL(2,\mathbb{C})$ space which
reproduces space-time twice \cite{key-15} . This suggests that the
non-locality of quantum mechanics could be an artifact created by
describing quantum amplitudes relative to an $SO(1,3)$ base space.
The idea of a direct connection between the complex structures of
quantum mechanics and the Lorentz group is not a new one \cite{key-16},
but has mainly been studied in the context of null structures. In
this paper we adapt canonical quantization to a spinorial space which
forms a substructure of Minkowski space and avoid the restriction
to null structures by means of Clifford algebras.

As shown in \cite{key-2}, the space-time coordinates $x^{\mu}$ of
four-dimensional Minkowski space can be resolved into Weyl spinors
according to
\begin{gather}
x^{A\dot{B}}=c^{A}\bullet c^{\ast\dot{B}},\quad c^{A}\bullet c^{B}=0,\quad x^{A\dot{B}}\stackrel{def}{=}\sigma_{\mu}^{A\dot{B}}x^{\mu},\quad u\bullet v\stackrel{def}{=}\frac{1}{2}\{u,v\},\label{eq:xcc}
\end{gather}
where the $\sigma$'s are the Hermitian Pauli matrices \cite{key-3},\cite{key-4}
and the components of $c^{A}$ belong to the complexified generating
space of the split Clifford algebra $\mathcal{C}l(4,4,\mathbb{R})$.
The product $\bullet$ is the inner product of the Clifford algebra.
A similar factorization is well known from the factorization of the
Lorentz metric $\eta_{\mu\nu}=\frac{1}{2}\{\gamma_{\mu},\gamma_{\nu}\}=\gamma_{\mu}\bullet\gamma_{\nu}$,
where the $\gamma$'s generate the Clifford algebra $\mathcal{C}l(1,3,\mathbb{R}$).
In general, an even-dimensional real Clifford algebra can be written
in complex form by a decomposition of the complexified generating
space (a \textit{polarization}) \cite{key-5}. This turns the algebra
into a dual system, in our case exemplified by $c$ and $c^{\star}$,
and is well known from the Clifford algebra of creation and annihilation
operators for a system of fermions. The dotted and undotted capital
letters refer to the transformation properties of the Weyl spinors
under $SL(2,\mathbb{C})$. 

Complex Weyl spinors can only generate a four-dimensional Minkowski
space, but there are reasons to believe that four dimensions do not
suffice to accommodate the symmetries of the Standard Model. Within
the framework of Weyl spinors, our only option of increasing the dimension
is to replace the complex numbers with a higher dimensional normed
division algebra. It has been conjectured that there is a connection
between the octonions and ten-dimensional Minkowski space \cite{key-7,key-8,key-9,key-10}.
One of the objects of this paper is to create a model which exemplifies
this connection.

The octonions \cite{key-7} $\mathcal{O}$ is a non-commutative normed
division algebra which is \textit{alternative}, but not \textit{associative}.
\textit{Alternative} means that the product of three numbers is associative
if at least two of them differ by only a real factor. An octonion
$z$ can be written as
\begin{gather*}
z=x_{0}e_{0}+\sum_{i=1}^{7}x_{i}e_{i},\quad x_{i}\in\mathbb{R},\quad e_{0}=1,\:(e_{i})^{2}=-1,\\
z^{\ast}\stackrel{def}{=}x_{0}e_{0}-\sum_{i=1}^{7}x_{i}e_{i},\quad(z_{1}z_{2})^{*}=z_{2}^{\ast}z_{1}^{\ast},
\end{gather*}
where $e_{i},\:i=1,\ldots,7.$ are the seven anti-commuting imaginary
units.

Consider the tensor product
\[
\boldsymbol{T}=\mathcal{\mathcal{O}\otimes_{\mathbb{R}}C}l(n,n,\mathbb{R})
\]
of the two $\mathbb{R}$-algebras $\mathcal{O}$ and $\mathcal{C}l(n,n,\mathbb{R})$.
Since $\mathcal{O}$ is non-commutative and non-associative, and $\mathcal{C}l(n,n,\mathbb{R})$
is non-commutative and associative, their tensor product will be a
non-commutative, non-associative $\mathbb{R}$-algebra. The octonionic
conjugation $\ast$ is trivially extended to $T$ by
\[
(z\otimes u)^{\ast}\stackrel{def}{=}z^{\ast}\otimes u,
\]
and the inner product of $\mathcal{C}l(n,n,\mathbb{R})$ defines the
non-commutative product in $\boldsymbol{T}$ 
\begin{gather}
(z_{1}\otimes u_{1})\times(z_{2}\otimes u_{2})\stackrel{def}{=}(z_{1}z_{2})\otimes(u_{1}\bullet u_{2}),\label{eq:define inner product}
\end{gather}
with the conjugation
\begin{equation}
(v_{i}\times v_{j})^{\ast}=v_{j}^{\ast}\times v_{i}^{\ast},\:v_{i},v_{j}\in\boldsymbol{T}.\label{eq:conjugation}
\end{equation}
When $u_{1}$ and $u_{2}$ belong to the generating space of $\mathcal{C}l(n,n,\mathbb{R})$,
then $u_{1}\bullet u_{2}=r\underline{1},\,r\in\mathbb{R}$ and  (\ref{eq:define inner product})
becomes
\begin{gather}
(z_{1}\otimes u_{1})\times(z_{2}\otimes u_{2})=(z_{1}z_{2}r)\otimes\underline{1}.\label{eq:real inner product}
\end{gather}
In this case we may leave out $\otimes\underline{1}$ and regard  (\ref{eq:real inner product})
as the octonion $z_{1}z_{2}r$. 

To extend the factorization (\ref{eq:xcc}) to the octonion case,
we consider the $n\times n$ matrix
\begin{gather}
H_{ij}=v_{i}\times v_{j}^{\ast},\:v_{i}\in\boldsymbol{T},\:i,j=1,n,\label{eq:octonion hermitian matrix}
\end{gather}
which according to  (\ref{eq:conjugation}) is identically Hermitian.
If the Clifford components of $v_{i}$ and  $v_{j}$ belong to the
generating space of $\mathcal{C}l(n,n,\mathbb{R})$ so that their
inner products are real, $H$ becomes an octonionic Hermitian matrix.
The following proposition shows that any $n\times n$ octonionic Hermitian
matrix can be obtained in this manner:
\begin{quote}
\textit{Proof. Consider the n vectors
\begin{gather*}
v_{i}=\sum_{k=1}^{n}a_{ik}\otimes\mathbf{e}_{k}+b_{ik}\otimes\mathbf{f}_{k},\:i=1,n.\\
\{\mathbf{e}_{i},\mathbf{e}_{j}\}=2\delta_{ij},\;\{\mathbf{f}_{i},\mathbf{f}_{j}\}=-2\delta_{ij},\:\{\mathbf{e}_{i},\mathbf{f}_{j}\}=0,\:i,j=1,n.
\end{gather*}
where $a_{ik}$ and $b_{ik}$ are octonions. Taking advantage of the
Hermitian symmetry of Equation (\ref{eq:octonion hermitian matrix}),
and setting $a_{ij}=b_{ij}=0$ for $i<j$, Equation (\ref{eq:octonion hermitian matrix})
can be written as the system of equations
\[
H_{11}=|a_{11}|^{2}-|b_{11}|^{2},\quad H_{1j}=a_{11}a_{j1}^{\ast}-b_{11}b_{j1}^{\ast},\,j=2,\ldots n,
\]
\begin{gather}
H_{ii}=|a_{ii}|^{2}-|b_{ii}|^{2}+\sum_{k=1}^{i-1}|a_{ik}|^{2}-|b_{ik}|^{2},\label{eq:proof octonion 1}\\
H_{ij}=a_{ii}a_{ji}^{\ast}-b_{ii}b_{ji}^{\ast}+\sum_{k=1}^{i-1}a_{ik}a_{jk}^{\ast}-b_{ik}b_{jk}^{\ast},\,j=i+1,\ldots n,\label{eq:octonion 3}
\end{gather}
for $i=2,\ldots n$. We proceed by induction. For an arbitrarily chosen
value of $i$, we assume that the terms after the summation sign in
 (\ref{eq:proof octonion 1}) do not depend on $a_{ii}$ and $b_{ii}$.
For $i=1$, this is trivially true since these terms are absent. Consequently,
Equation (\ref{eq:proof octonion 1}) can be solved with respect to
$a_{ii}$ and $b_{ii}$ so that neither of them vanishes. This solution
does not depend on $a_{ji}$ and $b_{ji},\,j=i+1,\ldots n.$, and
Equation (\ref{eq:octonion 3}) can therefore be solved with respect
to either of them. This solution does not depend on $a_{i+1\,i+1}$
and $b_{i+1\,i+1}$, and neither therefore do the terms after the
summation sign in Equation (\ref{eq:proof octonion 1}) for $H_{i+1\,i+1}$
depend on them.}
\end{quote}
\textit{\emph{As a corollary, it is easily verified that any $n\times n$
$\mathit{complex}$ Hermitian matrix can be factorized in terms of}}
$\mathcal{C}l(2n,2n,\mathbb{R})$ spinors\textit{\emph{ with the supplementary
condition $v_{i}\times v_{j}=0$. A similar result was obtained in
\cite{key-2} by means of a unitary similarity transformation and
is exemplified in Equation (\ref{eq:xcc}) for $n=2$. }}Applying
the above proposition to a ten-dimensional Minkowski space, we get
\begin{gather}
x^{A\dot{B}}=c^{A}\times c^{\ast\dot{B}},\quad x^{A\dot{B}}\stackrel{def}{=}\sigma_{\mu}^{A\dot{B}}x^{\mu},\quad c^{A}=\sum_{k=1}^{2}a_{k}^{A}\otimes\mathbf{e}_{k}+b_{k}^{A}\otimes\mathbf{f}_{k},\label{eq:xcc octonion}
\end{gather}
where $\sigma_{\mu}$ are the ten octonionic Hermitian Pauli matrices. 

The determinant of an octonionic Hermitian matrix is well-defined
\begin{gather*}
det\,\begin{pmatrix}a & c\\
c^{\ast} & b
\end{pmatrix}\stackrel{def}{=}ab-cc^{\ast},
\end{gather*}
and as in four-dimensional Minkowski space, we have $det(X)=x_{\mu}x^{\mu}$,
where $X$ is the octonionic Hermitian matrix with entries $x^{A\dot{B}}$. 

The octonionic spinors fall into two classes corresponding to the
Lorentz transformations
\[
\chi^{A}\rightarrow S_{\;\:B}^{A}\chi^{B},\quad\psi_{\dot{A}}\rightarrow-S_{\dot{A}}^{\ast\;\:\dot{B}}\psi_{\dot{B}},\quad S_{A}^{\;\:B}\stackrel{def}{=}\epsilon^{BE}S_{\;\:E}^{F}\epsilon_{FA},
\]
where $S$ contains only octonions from a single complex subspace
of $\mathcal{O}$. In case both $S$, $\chi^{A}$ and $\psi_{A}$
belong to the same complex subspace, the Lorentz transformations of
$\chi^{A}$ and $\epsilon^{AE}\psi_{E}^{\ast}$ are the same. As we
shall see, spinors transforming like $\chi^{A}$ and $\psi_{\dot{A}}$
generate coordinates and momenta respectively. In the appendix we
prove that 
\begin{gather}
(-\psi_{F}^{\ast}\,S_{A}^{\;\:F})(S_{\;\:E}^{A}\,\chi^{E})+o.c.=det(S)\,\psi_{A}^{\ast}\chi^{A}+o.c.,\quad det(S)\in\mathbb{R},\label{eq:lorentz invariance contraction}
\end{gather}
and that accordingly $Re(\psi_{A}^{\ast}\chi^{A})$ is Lorentz invariant
when $det(S)=1$. Since the basic types of $SO(1.9)$ transformations
can be obtained from a single transformation with $det(S)=1$, or
from two consecutive ones with $det(S)=-1$ \cite{key-11}, it follows
that $Re(\psi_{A}^{\ast}\chi^{A})$ is generally Lorentz invariant.
For two elements in $T$ of the form
\begin{equation}
c^{A}=\sum_{i}\chi_{i}^{A}\otimes u_{i},\quad d_{\dot{A}}=\sum_{j}\psi_{\dot{Aj}}\otimes v_{j},\quad u_{i},v_{j}\in V,\label{eq:coordinatemomentumdefined}
\end{equation}
where $V$ is the generating space of $\mathcal{C}l(n,n,\mathbb{R})$,
it follows that $Re(d_{A}^{\ast}\times c^{A})$ is Lorentz invariant.

Under a Lorentz transformation of $c^{A}$, the coefficients in Equation
(\ref{eq:xcc octonion}) transform according to
\begin{gather*}
a_{k}^{A}\rightarrow S_{\;\:B}^{A}a_{k}^{B},\quad b_{k}^{A}\rightarrow S_{\;\:B}^{A}b_{k}^{B},\;\text{or}\;a_{k}\rightarrow S\,a_{k},\quad b_{k}\rightarrow S\,b_{k},
\end{gather*}
which makes
\begin{gather*}
x^{A\dot{B}}=\sum_{k=1}^{2}a_{k}^{A}a_{k}^{\ast\dot{B}}-b_{k}^{A}b_{k}^{\ast\dot{B}}
\end{gather*}
transform as
\begin{gather}
X\rightarrow X^{'}=\sum_{k=1}^{2}(S\,a_{k})(S\,a_{k})^{\dagger}-(S\,b_{k})(S\,b_{k})^{\dagger}.\label{eq:XsXsab-1}
\end{gather}
Manogue and Dray \cite{key-11} found that the compatibility condition
\begin{gather*}
(Sv)(Sv)^{\dagger}=S(v\,v^{\dagger})S^{\dagger}
\end{gather*}
between the spinor and vector representations is satisfied iff $S$
contains only octonions from a single complex subspace and $det(S)\in\mathbb{R}$.
If therefore we assume that $det(S)=\pm1$, Equation (\ref{eq:XsXsab-1})
becomes
\begin{gather}
X\rightarrow X^{'}=S\,X\,S^{\dagger}.\label{eq:XsXs}
\end{gather}
There is no associativity ambiguity because the octonions are alternative
and only one complex subspace is used in each transformation. The
determinant of $X^{'}$ is \cite{key-11}
\[
det(X^{'})=det(S\,S^{\dagger})\,det(X),
\]
and since $det(S\,S^{\dagger})=det(S)\,det(S)^{*}=1$, it follows
that $det(X)$ is preserved and that Equation (\ref{eq:XsXs}) therefore
generates a Lorentz transformation of $x^{\mu}$. Arbitrary finite
Lorentz transformations are obtained by consecutive (nested) application
of transformations corresponding to different complex subspaces \cite{key-8}.

\section{The classical point particle}

We use the Hamiltonian, rather than the Lagrangian formulation of
mechanics, as our starting point. It is better suited to a pure spinor
model and is a direct precursor to quantum mechanics. It is also more
compatible with an octonionic generalization. The kinetic momentum
constraint, which in the Lagrangian formulation is a consequence of
reparametrization invariance, becomes a constraint in the Hamiltonian
formulation, and is imposed by means of a Lagrange multiplier (the
einbein). In section 4 we show that the kinetic momentum constraint
for a point particle follows from the equations of motion for the
metric and its associated scalar field in the point particle limit
of the spinor string. 

The Point particle in spinor space will be described by spinor coordinates
and momenta $c^{A}(\tau)$ and $d_{\dot{A}}(\tau)$ of the form (\ref{eq:coordinatemomentumdefined}),
where $\tau$ is a parameter time. They determine the space-time coordinates
and momenta through 
\begin{equation}
x^{A\dot{B}}\stackrel{def}{=}c^{A}\times c^{\ast\dot{B}},\quad p_{A\dot{B}}\stackrel{def}{=}d_{\dot{B}}\times d_{A}^{\ast}.
\end{equation}
When $det(S)\in\mathbb{R}$, then $det(-S_{\dot{A}}^{\ast\;\:\dot{B}})=det(S)$,
and a Lorentz transformation of $d_{\dot{B}}$ makes $p_{A\dot{B}}$
transform like $X$ in Equation (\ref{eq:XsXs}) with $S$ and $S^{\dagger}$
switched around. The Hamiltonian action takes the form
\begin{gather}
I=\int d\tau\bigl(\sqrt{lm}\,d_{A}^{\ast}\times\frac{dc^{A}}{d\tau}+o.c.-le(\tau)\mathcal{H}(x^{\mu},p{}^{\nu})\bigr),\label{eq:point particle action}
\end{gather}
where $\mathcal{H}(x^{\mu},p{}^{\nu})$ is the Hamiltonian associated
with the kinetic momentum constrain, and $e(\tau)$ is an einbein.
The first two terms in the Lagrangian are the real part of an octonionic
spinor contraction which, according to Equation (\ref{eq:lorentz invariance contraction}),
is Lorentz invariant. $l$ and $m$ are constants with the dimension
of length and mass respectively. 

We shall return to the octonion model in the end of section 4, but
in the following we shall restrict ourselves to the complex case with
a four-dimensional Minkowski space. The product $\times$ of $T$
is hereby replaced by the inner product $\bullet$ of $\mathcal{C}l(4,4,\mathbb{R})$
written in complex form. Differentiation with respect to spinor-valued
variables is defined through
\begin{gather*}
\delta f=\frac{1}{2}\{\partial f/\partial c^{A},\delta c^{A}\},
\end{gather*}
and leads to the differentiation rules
\[
\partial c^{A}/\partial c^{B}=\delta_{B}^{A},\quad\partial(d_{A}^{\ast}\bullet c^{A})/\partial c^{B}=d_{B}^{\ast},\quad\partial f(x)/\partial c^{A}=c^{\ast\dot{B}}\partial f(x)/\partial x^{A\dot{B}}.
\]
The Poisson bracket in spinor space can then be defined as the `Clifford
bracket'
\begin{multline*}
\left\{ N,M\right\} _{C.B.}\stackrel{def}{=}\frac{1}{2\sqrt{lm}}\bigl(\{\partial N/\partial c^{A},\partial M/\partial d_{A}^{\ast}\}+\{\partial N/\partial c^{\ast\dot{A}},\partial M/\partial d_{\dot{A}}\}\\
-\{\partial M/\partial c^{A},\partial N/\partial d_{A}^{\ast}\}-\{\partial M/\partial c^{\ast\dot{A}},\partial N/\partial d_{\dot{A}}\}\bigr),
\end{multline*}
which is skew-symmetric in $N$ and $M$ and real when $N$ and $M$
are real. By means of this bracket, the equations of motion obtained
from the action (\ref{eq:point particle action}) by independent variation
of $c$ and $d$ can be written as
\begin{gather}
\frac{d}{d\tau}c^{A}=le(\tau)\left\{ c^{A},\mathcal{H}\right\} _{C.B.},\quad\frac{d}{d\tau}d_{A}^{\ast}=le(\tau)\left\{ d_{A}^{\ast},\mathcal{H}\right\} _{C.B.},\label{eq:eqs of motion c,d}\\
\frac{d}{d\tau}x^{\mu}=le(\tau)\left\{ x^{\mu},\mathcal{H}\right\} _{C.B.},\quad\frac{d}{d\tau}p_{\mu}=le(\tau)\left\{ p_{\mu},\mathcal{H}\right\} _{C.B.}.\label{eq:eqs of motion x,p}
\end{gather}
The action (\ref{eq:point particle action}) has a global $SL(2,\mathbb{C})$
and $U(1)$ gauge symmetry with the conserved Noether charges
\begin{gather*}
\mathcal{J_{AB}}=d_{A}^{\ast}\bullet c_{B}+d_{B}^{\ast}\bullet c_{A},\quad\jmath=i(d_{A}^{\ast}\bullet c^{A}-d_{\dot{A}}\bullet c^{\ast\dot{A}}).
\end{gather*}
To obtain a pure space-time system where the equations of motion (\ref{eq:eqs of motion x,p})
contain only $x$ and $p$ themselves, we must require that they vanish
\begin{gather*}
d_{A}^{\ast}\bullet c_{B}+d_{B}^{\ast}\bullet c_{A}=0,\quad d_{A}^{\ast}\bullet c^{A}-d_{\dot{A}}\bullet c^{\ast\dot{A}}=0,
\end{gather*}
which is equivalent to
\begin{gather}
d_{A}^{\ast}\bullet c^{B}=\mu(\tau)\,\delta_{A}^{B},\quad\mu(\tau)\in\mathbb{R}.\label{eq:d cequal to 1}
\end{gather}
In this case the Clifford bracket becomes proportional to the ordinary
Poisson bracket: 
\begin{multline}
\left\{ N(x,p),M(x,p)\right\} _{C.B.}=\frac{1}{\sqrt{lm}}\Bigl(\frac{\partial N}{\partial x^{\mu}}\frac{\partial M}{\partial p_{\nu}}-\frac{\partial M}{\partial x^{\mu}}\frac{\partial N}{\partial p_{\nu}}\Bigr)\Bigl(\frac{1}{8}\sigma_{A\dot{B}}^{\mu}\sigma_{\nu}^{A\dot{F}}\{c^{\ast\dot{B}},d_{\dot{F}}\}\\
+c.c.\Bigr)=\frac{\mu}{\sqrt{lm}}\Bigl(\frac{\partial N}{\partial x^{\mu}}\frac{\partial M}{\partial p_{\mu}}-\frac{\partial M}{\partial x^{\mu}}\frac{\partial N}{\partial p_{\mu}}\Bigr)=\frac{\mu}{\sqrt{lm}}\left\{ N(x,p),M(x,p)\right\} _{P.B.}.\label{eq:proportionality clifford poisson}
\end{multline}
When $\mu(\tau)\neq0$, the reparametrization
\begin{gather}
\frac{d\overline{\tau}}{d\tau}=\sqrt{\frac{l}{m}}e(\tau)\mu(\tau),\label{eq:reparametrization}
\end{gather}
turns the equations of motion (\ref{eq:eqs of motion x,p}) into the
usual space-time canonical equations of motion with proper time $\overline{\tau}$
\begin{gather}
\frac{d}{d\overline{\tau}}x^{\mu}=\left\{ x^{\mu},\mathcal{H}(x,p\right\} _{P.B.},\quad\frac{d}{d\overline{\tau}}p_{\mu}=\left\{ p_{\mu},\mathcal{H}(x,p)\right\} _{P.B.}\label{eq:eqs of motion forxp}
\end{gather}
and `hides' the spinor substructure. For a free particle with the
parametrization $e(\tau)=1$, proper time becomes
\begin{gather*}
\overline{\tau}=\frac{lM}{4m}(\tau-\tau_{0})^{2}+\overline{\tau}_{0},
\end{gather*}
and consequently, the spinor paths $c(\tau)$,$d(\tau)$ reproduce
the space-time paths $x(\overline{\tau}),$$p(\overline{\tau})$ twice. 

\section{The quantum point particle}

As a first step towards quantization, we shall describe the motion
of the classical point particle by $N$ integral curves in spinor
coordinate-momentum space and show that they form a unitarily invariant
system. Let $c_{i}^{A}(\tau),d_{iA}^{*}(\tau),\,i=1\ldots,N$ be $N$
solutions to the equations of motion (\ref{eq:eqs of motion c,d})
and let us assemble them into two $N$-dimensional ket and bra vectors
$\stackrel{>}{C}$ and $\stackrel{<}{D}$. We shall assume that their
components belong to the complexified generating space of  $Cl(4N,4N,\mathbb{R})$
and that all inner products between coordinates and momenta on different
integral curves vanish. This means that the two $N\times N$ Hermitian
matrices
\begin{gather}
X^{A\dot{B}}=\stackrel{>}{C^{A}}\bullet\stackrel{<}{C^{\dot{B}}},\quad P_{A\dot{B}}=\stackrel{>}{D_{\dot{B}}}\bullet\stackrel{<}{D_{A}},\label{eq:XABdef}
\end{gather}
are diagonal. The diagonal entries are the space-time coordinates
and momenta of the integral curves. Since $X$ and $P$ are diagonal,
they trivially commute with each other
\begin{gather}
[X^{\mu},X^{\nu}]=[P_{\mu},P_{\nu}]=[X^{\mu},P_{\nu}]=0.\label{eq:commutator X,P}
\end{gather}
The Noether charge condition (\ref{eq:d cequal to 1}) can be written
as
\begin{gather}
\stackrel{>}{C^{A}}\bullet\stackrel{<}{D_{B}}=\mu(\tau)\,\delta_{B}^{A}\cdot\underline{1},\quad\mu(\tau)\in\mathbb{R}.\label{eq:noether condition C D}
\end{gather}
The $N$ integral curves may be regarded as a single solution $\stackrel{>}{C},\stackrel{<}{D}$
to the equations of motion for an action describing $N$ independent
identical point particles. If the Hamiltonian is a polynomial expression
in $x$ and $p$, this action can be written as
\begin{gather}
\int d\tau\,Tr\,\Bigl(\sqrt{lm}\frac{d}{d\tau}\stackrel{>}{C^{A}}\bullet\stackrel{<}{D_{A}}+h.c.-le(\tau)H(X,P)\Bigr).\label{eq:action for N parrticles}
\end{gather}
The system (\ref{eq:commutator X,P}), (\ref{eq:noether condition C D})
and (\ref{eq:action for N parrticles}) is preserved by the global
$U(N)$ transformations
\begin{gather*}
\stackrel{>}{C^{A}}\rightarrow U\stackrel{>}{\,C^{A},}\quad\stackrel{<}{D_{A}}\rightarrow\stackrel{<}{D_{A}\,}U^{\dagger},\\
X^{\mu}\rightarrow UX^{\mu}U^{\dagger},\quad P_{\mu}\rightarrow UP_{\mu}U^{\dagger},
\end{gather*}
which produce artificial couplings between different integral curves.
Conversely, given this unitarily invariant system, the $N$ integral
curves are recovered when $X$ and $P$ are diagonalized by a global
unitary similarity transformation. The appearance of a $U(N)$ symmetry
should not come as a surprise since it is already present in the complex
form of the generating algebra of $Cl(4N,4N,\mathbb{R})$.

The observables of the system are the coordinates and momenta of the
original $N$ integral curves. They can be characterized in a unitarily
invariant manner as the eigenvalues of $X$ and $P$. If the point
particle is restricted to move along one of the integral curves, we
can define the `state' of the particle as the integral curve along
which it is moving. This information can be encoded in a `state vector'
$|\,s>$ which is an eigenvector of $X^{\mu}$ and is constant in
time. The spinor coordinate corresponding to an eigenvalue can be
written as an `expectation value'
\begin{gather}
E(\stackrel{>}{C^{A}})\stackrel{def}{=}<s\,|\,\stackrel{>}{C^{A}},\label{eq:expectation value}
\end{gather}
which is unitarily invariant since $<s\,|$ transforms like a bra
vector. This `expectation value' will, in time, move through the spinor
coordinates corresponding to the space-time eigenvalues on the integral
curve selected by $<s\,|$. 

To quantize the classical point particle system, we must translate
the Clifford brackets $\left\{ c^{A},M(x,p)\right\} _{C.B.}$ and
$\left\{ d_{A}^{*},M(x,p)\right\} _{C.B.}$ in the equations of motion
(\ref{eq:eqs of motion c,d}), in such a manner that (\ref{eq:eqs of motion forxp})
becomes the Heisenberg equations of motion for $X$ and $P$. This
is accomplished by
\begin{gather}
c^{A}\rightarrow\stackrel{>}{C^{A}},\;d_{A}^{*}\rightarrow\stackrel{<}{D_{A}},\:x^{A\dot{B}}\rightarrow\stackrel{>}{C^{A}}\bullet\stackrel{<}{C^{\dot{B}}},\;p_{A\dot{B}}\rightarrow\stackrel{>}{D_{\dot{B}}}\bullet\stackrel{<}{D_{A}},\nonumber \\
\left\{ c^{A},H(x,p)\right\} _{C.B.}\rightarrow\frac{1}{2i\hslash}\frac{1}{\sqrt{lm}}[X^{A\dot{B}},H(X,P)]\stackrel{>}{D_{\dot{B}}},\nonumber \\
\left\{ d_{A}^{*},H(x,p)\right\} _{C.B.}\rightarrow\frac{1}{2i\hslash}\frac{1}{\sqrt{lm}}\stackrel{<}{C^{\dot{B}}}[P_{A\dot{B}},H(X,P)],\label{eq:quantization rule 1}
\end{gather}
which turns (\ref{eq:eqs of motion c,d}) into
\begin{gather*}
\frac{d}{d\tau}\stackrel{>}{C^{A}}=\frac{1}{2i\hslash}e(\tau)\sqrt{\frac{l}{m}}[X^{A\dot{B}},H(X,P)]\stackrel{>}{D_{\dot{B}}},\\
\frac{d}{d\tau}\stackrel{<}{D_{A}}=\frac{1}{2i\hslash}e(\tau)\sqrt{\frac{l}{m}}\stackrel{<}{C^{\dot{B}}}[P_{A\dot{B}},H(X,P)],
\end{gather*}
and which by use of (\ref{eq:XABdef}) and the Noether charge condition
(\ref{eq:noether condition C D}) yields
\begin{gather*}
\frac{d}{d\tau}X^{A\dot{B}}=\frac{1}{i\hslash}\sqrt{\frac{l}{m}}e(\tau)\mu(\tau)[X^{A\dot{B}},H(X,P)],\\
\frac{d}{d\tau}P_{A\dot{B}}=\frac{1}{i\hslash}\sqrt{\frac{l}{m}}e(\tau)\mu(\tau)[P_{A\dot{B}},H(X,P)].
\end{gather*}
After a reparametrization (\ref{eq:reparametrization}) they become
the Heisenberg equations of motion.

Applying the quantization rule (\ref{eq:quantization rule 1}) to
the Clifford bracket $\{x,p\}$ gives
\begin{gather*}
\bigl\{ x^{A\dot{B}},p_{G\dot{F}}\bigl\}_{C.B.}=c^{A}\bullet\bigl\{ c^{\ast\dot{B}},p_{G\dot{F}}\bigr\}_{C.B.}+\bigl\{ c^{A},p_{G\dot{F}}\bigl\}_{C.B.}\bullet\,c^{\ast\dot{B}}\\
\rightarrow\frac{1}{2i\hslash\sqrt{lm}}\stackrel{>}{\bigl(C^{A}}\bullet\stackrel{<}{D_{E}}[X^{E\dot{B}},P_{G\dot{F}}]+[X^{A\dot{E}},P_{G\dot{F}}]\stackrel{>}{D_{\dot{E}}}\bullet\stackrel{<}{C^{\dot{B}}}),
\end{gather*}
which by use of the Noether charge condition (\ref{eq:noether condition C D})
becomes
\[
\bigl\{ x^{A\dot{B}},p_{G\dot{F}}\bigl\}_{C.B.}\rightarrow\frac{\mu}{i\hslash\sqrt{lm}}[X^{A\dot{B}},P_{G\dot{F}}],
\]
and from the proportionality (\ref{eq:proportionality clifford poisson})
between Clifford and Poisson brackets, gives the usual quantum condition
\[
\bigr\{ x^{A\dot{B}},p_{G\dot{F}}\bigl\}_{P.B.}\rightarrow\frac{1}{i\hslash}\left[X^{A\dot{B}},P_{G\dot{F}}\right].
\]

In the quantum matrix system, X and P are conjugate variables and,
consequently, can no longer be turned into a set of integral curves
through diagonalization. The state vector which used to describe the
state of the system as a choice of integral curve now takes on a more
abstract character. In both the classical and the quantum system the
state vector is constant in time. But in the quantum system, after
a measurement has been performed, the expectation value (\ref{eq:expectation value})
will not remain a spinor coordinate corresponding to an eigenvalue,
but evolve in time into a complex linear combination of spinor coordinates
corresponding to different eigenvalues. The measurements are no longer
predictable, but become stochastic in accordance with the Born rule.
In the non-relativistic limit the time evolution of the state vector
can be turned into the matrix form of the Schrödinger equation by
a local unitary transformation (the Schrödinger picture) \cite{key-2}.
Regardless of its abstract character, the formal resemblance of the
quantum mechanical wave function to the classical state vector lends
support to the view that it is an information-carrying object rather
than a primary physical variable.

\section{The classical string}

The quantized relativistic point particle is a valid concept only
in the approximation when pair production can be ignored. As a direct
generalization, we consider a string carrying the spinor fields $c^{A}(\tau,\sigma)$
and $d_{\dot{A}}(\tau,\sigma)$ which are two-component spinors, both
in relation to space-time and in relation to the worldsheet. The worldsheet
spinor indices will be suppressed. The spinor components belong to
the complexified generating space of an infinite-dimensional split
Clifford algebra. We follow the convention that $\mu,\nu,\ldots$
denote the space-time indices and $\alpha,\beta,\ldots$, the worldsheet
indices. Differentiation with respect to the worldsheet parameters
$\sigma^{\alpha}=\tau,\sigma$ is written as $\partial_{\alpha}$.
The spinors $c^{A}$ and $d_{\dot{A}}$ determine the space-time coordinates
and the space-time momentum current according to
\begin{gather}
x^{A\dot{B}}=\overline{c}^{\dot{B}}\bullet c^{A},\nonumber \\
p_{A\dot{B}}^{a}=\overline{d}_{A}\bullet\rho^{a}d_{\dot{B}},\label{eq:mom cur density}\\
\rho^{0}\stackrel{def}{=}\begin{pmatrix}0 & 1\\
1 & 0
\end{pmatrix},\;\rho^{1}\stackrel{def}{=}\begin{pmatrix}0 & -1\\
1 & 0
\end{pmatrix},\quad\{\rho^{a},\rho^{b}\}=2\eta^{ab},\quad\eta^{ab}\stackrel{def}{=}diag(1,-1),\nonumber \\
\overline{\psi}\stackrel{def}{=}\psi^{\dagger}\rho^{0},\quad\overline{\rho^{a}}\stackrel{def}{=}\rho^{0}(\rho^{a})^{\dagger}\rho^{0}=\rho^{a}.\nonumber 
\end{gather}
$\rho^{0}$ and $\rho^{1}$ are the Dirac matrices in $1+1$ dimensions
and $x^{\mu}$ and $p_{\mu}^{a}$ are worldsheet scalars and vectors
respectively. 

For a string residing in space-time, the Lorentz metric $\eta_{\mu\nu}$
induces a metric on the worldsheet through the tangent derivatives
$\partial_{\alpha}x^{\mu}$. In spinor space, these are replaced by
the complex vectors
\begin{gather*}
V_{\alpha}^{\mu}\stackrel{def}{=}\sigma_{A\dot{B}}^{\mu}\partial_{\alpha}\overline{c}^{\dot{B}}\bullet c^{A}
\end{gather*}
with the real part $\partial_{\alpha}x^{\mu}$. The Hermitian tensor
\begin{gather*}
g_{\alpha\beta}\stackrel{def}{=}V_{\alpha}^{\mu}V_{\beta}^{\nu\ast}\eta_{\mu\nu},\quad g_{\alpha\beta}^{\ast}=g_{\beta\alpha}
\end{gather*}
can be split into a metric $h_{\alpha\beta}$ and a scalar field $\phi$.
The metric allows the space-time momentum current (\ref{eq:mom cur density})
to be written in covariant form as a current density
\begin{gather}
p_{A\dot{B}}^{\alpha}=e\,\overline{d}_{A}\bullet\rho^{\alpha}d_{\dot{B}},\label{eq:covarian sp_t_mome_density}\\
h_{\alpha\beta}=e_{\alpha}^{\;\:a}e_{\beta}^{\;\:b}\eta_{ab},\quad e\stackrel{def}{=}det(e_{\alpha}^{\;\:a}),\quad\rho^{\alpha}\stackrel{def}{=}e_{\;\:a}^{\alpha}\rho^{a}.\nonumber 
\end{gather}

The simplest string action is
\begin{gather}
\int d\tau d\sigma\,e\Bigl(\sqrt{lm}\;\overline{d_{A}}\bullet\rho^{\alpha}\nabla_{\alpha}c^{A}+c.c.-\frac{l}{m}(\overline{d}_{A}\bullet d_{\dot{B}})(\overline{d}^{A}\bullet d^{\dot{B}})+lm\phi R(h_{\alpha\beta})\Bigr),\label{eq:stringaction}
\end{gather}
where $l$ and $m$ are constants with dimension of length and mass
respectively. The equations of motion obtained by independent variation
of $c^{A}$, $d_{\dot{B}}$, $\phi$ and $e_{\;\alpha}^{a}$  are
\begin{gather}
\rho^{\alpha}\nabla_{\alpha}c^{A}=\frac{2\sqrt{lm}}{m^{2}}(\overline{d}^{A}\bullet d^{\dot{B}})d_{\dot{B}},\label{eq:c egs of motion}\\
\rho^{\alpha}\nabla_{\alpha}d_{\dot{A}}=0,\label{eq:d egs of motion}\\
2e_{\;\:\alpha}^{a}\nabla^{2}\phi-2e^{\beta a}\nabla_{\alpha}\nabla_{\beta}\phi+\partial L_{M}/\partial e_{\;\:a}^{\alpha}=0,\quad R(h_{\alpha\beta})=0,\nonumber 
\end{gather}
where the covariant derivative $\nabla_{\alpha}$ satisfies the `tetrad
postulate' 
\[
\nabla_{\alpha}e_{\;\:a}^{\beta}\stackrel{def}{=}\partial_{\alpha}e_{\;\:a}^{\beta}+\Gamma_{\gamma a}^{\beta}(h)e_{\;\:a}^{\gamma}-\omega_{\:a\alpha}^{b}e_{\;\:b}^{\beta}=0,
\]
with $\varGamma$ as a metric connection. Since the scalar curvature
vanishes, we can choose a parametrization in which $h_{\alpha\beta}=\eta_{\alpha\beta},\;e_{\;\:a}^{\alpha}=\delta_{a}^{\alpha}$
and $\nabla_{\alpha}=\partial_{\alpha}$. From Equation (\ref{eq:d egs of motion})
it follows that  the space-time momentum current density (\ref{eq:covarian sp_t_mome_density})
is conserved.

We shall find the class of solutions where $\overline{d}_{A}\bullet d_{\dot{B}}$
is constant and invertible (not a null vector). Applying $\rho^{\alpha}\partial_{\alpha}$
to both sides of Equation (\ref{eq:c egs of motion}) and using Equation
(\ref{eq:d egs of motion}), we find that $c^{A}$ satisfies the wave
equation and can be expanded according to
\begin{gather}
c^{A}=c^{A}(0)+l^{A}\tau+\sum_{n\neq0}a^{A}(n)e^{i\frac{1}{2}n(\tau+\sigma)}+b^{A}(n)e^{i\frac{1}{2}n(\tau-\sigma)},\:0\leq\sigma\leq\pi.\label{eq:c expansion}
\end{gather}
By means of the identity
\begin{gather}
H_{A\dot{E}}H^{B\dot{E}}\equiv\frac{1}{2}H_{F\dot{E}}H^{F\dot{E}}\delta_{A}^{B}\label{eq:identity}
\end{gather}
for Hermitian second rank spinors, Equation (\ref{eq:c egs of motion})
can be solved with respect to $d_{\dot{B}}$, giving
\begin{gather}
d_{\dot{B}}=k_{A\dot{B}}\rho^{\alpha}\partial_{\alpha}c^{A},\quad k_{A\dot{B}}\stackrel{def}{=}\frac{m^{2}}{\sqrt{lm}}\bigl(\overline{d}_{E}\bullet d_{\dot{F}}\,\overline{d}^{E}\bullet d^{\dot{F}}\bigr)^{-1}\overline{d}_{A}\bullet d_{\dot{B}},\label{eq:d}
\end{gather}
with the expansion
\begin{gather}
d_{\dot{B}}=k_{A\dot{B}}\rho^{\alpha}\bigl(l^{A}\delta_{\alpha}^{0}+\frac{i}{2}\sum_{n\neq0}na^{A}(n)k_{\alpha}^{L}e^{i\frac{1}{2}n(\tau+\sigma)}+nb^{A}(n)k_{\alpha}^{R}e^{i\frac{1}{2}n(\tau-\sigma)}\bigr),\label{eq:d expansion}\\
k_{\alpha}^{L}\stackrel{def}{=}(1,1),\quad k_{\alpha}^{R}\stackrel{def}{=}(1,-1).\nonumber 
\end{gather}
Since the Clifford algebra is infinite-dimensional, we can arrange
it so that the components of Fourier coefficients corresponding to
different modes are orthogonal to each other:
\begin{gather*}
a\bullet b^{\dagger}=0,\quad a(n)\bullet a^{\dagger}(m)=b(n)\bullet b^{\dagger}(m)=0\quad\text{for}\quad n\neq\pm m.
\end{gather*}
By using the fact that the wave vectors $k^{L}$ and $k^{R}$ are
null vectors, we get
\begin{equation}
\overline{d}_{A}\bullet d_{\dot{B}}=k_{A\dot{F}}\bigl(\overline{l}^{\dot{F}}\bullet l^{E}\bigr)k_{E\dot{B}},\label{eq:dd}
\end{equation}
which is constant in accordance with our ansatz. From Equations  (\ref{eq:d})
and (\ref{eq:dd}) it follows that
\begin{equation}
\overline{l}^{\dot{B}}\bullet l^{A}=\frac{2m^{2}}{\sqrt{lm}}(k_{F\dot{E}}k^{F\dot{E}})^{-2}k^{A\dot{B}}.\label{eq:l=00003Dk}
\end{equation}

The conserved $SL(2.\mathbb{C})$ Noether current density corresponding
to the string action (\ref{eq:stringaction}) is
\begin{gather*}
\jmath_{AB}^{\gamma}\stackrel{def}{=}\sqrt{lm}\,e\,(\overline{d}_{A}\bullet\rho^{\gamma}c_{B}+\overline{d}_{B}\bullet\rho^{\gamma}c_{A}),
\end{gather*}
which contains the angular momentum current density. From Equations
 (\ref{eq:c expansion}) and (\ref{eq:d expansion}), we obtain the
expansions of the space-time momentum and angular momentum current
densities
\begin{multline}
p_{A\dot{B}}^{\gamma}=k_{A\dot{F}}\Bigl(\overline{l}^{\dot{F}}\bullet\rho^{0}\rho^{\gamma}\rho^{0}l^{E}+\sum_{n\neq0}\frac{n^{2}}{4}\bigl(\overline{a}^{\dot{F}}(n)\bullet k_{\alpha}^{L}\rho^{\alpha}\rho^{\gamma}k_{\beta}^{L}\rho^{\beta}a^{E}(n)\\
+\overline{b}^{\dot{F}}(n)\bullet k_{\alpha}^{R}\rho^{\alpha}\rho^{\gamma}k_{\beta}^{R}\rho^{\beta}b^{E}(n)-\overline{a}^{\dot{F}}(n)\bullet k_{\alpha}^{L}\rho^{\alpha}\rho^{\gamma}k_{\beta}^{L}\rho^{\beta}a^{E}(-n)e^{-in(\tau+\sigma)}\\
-\overline{b}^{\dot{F}}(n)\bullet k_{\alpha}^{R}\rho^{\alpha}\rho^{\gamma}k_{\beta}^{R}\rho^{\beta}b^{E}(-n)e^{-in(\tau-\sigma)}\Bigr)k_{E\dot{B}},\label{eq:expansion p}
\end{multline}
\begin{multline}
\jmath_{AB}^{\gamma}=-\frac{i}{2}\sqrt{lm}\,k_{A\dot{F}}\sum_{n\neq0}n\overline{a}^{\dot{F}}(n)\bullet k_{\beta}^{L}\rho^{\beta}\rho^{\gamma}a_{B}(n)+n\overline{b}^{\dot{F}}(n)\bullet k_{\beta}^{R}\rho^{\beta}\rho^{\gamma}b_{B}(n)\\
+n\overline{a}^{\dot{F}}(n)\bullet k_{\beta}^{L}\rho^{\beta}\rho^{\gamma}a_{B}(-n)e^{-in(\tau+\sigma)}+n\overline{b}^{\dot{F}}(n)\bullet k_{\beta}^{R}\rho^{\beta}\rho^{\gamma}b_{B}(-n)e^{-in(\tau-\sigma)}\\
+A\leftrightarrow B\label{eq:expansion J}
\end{multline}
The boundary condition says that there is no flow of momentum and
angular momentum at the endpoints of the string
\begin{gather*}
p_{A\dot{B}}^{1}(\tau,0)=p_{A\dot{B}}^{1}(\tau,\pi)=0,\quad\jmath_{AB}^{1}(\tau,0)=\jmath_{AB}^{1}(\tau,\pi)=0,
\end{gather*}
which gives the relations between the $a$- and $b$ coefficients:
\begin{gather*}
\overline{a}^{\dot{B}}(n)\bullet\begin{pmatrix}0 & 0\\
1 & 0
\end{pmatrix}a^{A}(\pm n)=\overline{b}^{\dot{B}}(n)\bullet\begin{pmatrix}0 & 1\\
0 & 0
\end{pmatrix}b^{A}(\pm n),\\
\overline{a}^{\dot{B}}(n)\bullet\begin{pmatrix}0 & 0\\
0 & 1
\end{pmatrix}a^{A}(\pm n)=\overline{b}^{\dot{B}}(n)\bullet\begin{pmatrix}1 & 0\\
0 & 0
\end{pmatrix}b^{A}(\pm n),
\end{gather*}
and the vanishing of the constant momentum flow
\begin{gather}
\overline{l}^{\dot{F}}\bullet\rho^{1}l^{E}=0.\label{eq:lrho1vanish}
\end{gather}
This simplifies the expansions (\ref{eq:expansion p}) and (\ref{eq:expansion J})
into
\begin{gather}
p_{A\dot{B}}^{0}=k_{A\dot{F}}\biggl(\overline{l}^{\dot{F}}\bullet\rho^{0}l^{E}+\sum_{n\neq0}2n^{2}\overline{a}^{\dot{F}}(n)\bullet\begin{pmatrix}0 & 0\\
1 & 0
\end{pmatrix}a^{E}(n)\nonumber \\
-2n^{2}\overline{a}^{\dot{F}}(n)\bullet\begin{pmatrix}0 & 0\\
1 & 0
\end{pmatrix}a^{E}(-n)e^{-in\tau}cos(n\sigma)\biggl)k_{E\dot{B}},\label{eq:p0}\\
p_{A\dot{B}}^{1}=-k_{A\dot{F}}\biggl(\sum_{n\neq0}2in^{2}\overline{a}^{\dot{F}}(n)\bullet\begin{pmatrix}0 & 0\\
1 & 0
\end{pmatrix}a^{E}(-n)e^{-in\tau}sin(n\sigma)\biggl)k_{E\dot{B}},\nonumber \\
\jmath_{AB}^{0}=-2i\,\sqrt{lm}\,k_{A\dot{F}}\sum_{n\neq0}n\overline{a}^{\dot{F}}(n)\bullet\begin{pmatrix}0 & 0\\
0 & 1
\end{pmatrix}a_{B}(n)\nonumber \\
+n\overline{a}^{\dot{F}}(n)\bullet\begin{pmatrix}0 & 0\\
0 & 1
\end{pmatrix}a_{B}(-n)\,e^{-in\tau}cos(n\sigma)+A\leftrightarrow B,\label{eq:j0}\\
\jmath_{AB}^{1}=-2i\sqrt{lm}\,k_{A\dot{F}}\sum_{n\neq0}n\overline{a}^{\dot{F}}(n)\bullet\begin{pmatrix}0 & 0\\
0 & 1
\end{pmatrix}a_{B}(-n)\,e^{-in\tau}sin(n\sigma)+A\leftrightarrow B.\nonumber 
\end{gather}
The boundary condition $\partial_{\sigma}x=0$ gives
\[
\overline{a}^{\dot{B}}(n)\bullet a^{A}(\pm n)=\overline{b}^{\dot{B}}(n)\bullet b^{A}(\pm n),
\]
and turns the expansion of the space-time coordinates into a standing
wave
\begin{multline}
x^{A\dot{B}}\stackrel{def}{=}\overline{c}^{\dot{B}}\bullet c^{A}=\overline{c}^{\dot{B}}(0)\bullet c^{A}(0)+\overline{l}^{\dot{B}}\bullet l^{A}\tau^{2}\\
+2\sum_{n\neq0}\overline{a}^{\dot{B}}(n)\bullet a^{A}(n)+\overline{a}^{\dot{B}}(n)\bullet a^{A}(-n)e^{-in\tau}cos(n\sigma).\label{eq:space-time coord}
\end{multline}

We expect the total space-time momentum and the velocity of the center
of mass:
\begin{gather*}
p_{A\dot{B}}^{tot}\stackrel{def}{=}\int_{0}^{\pi}d\sigma p_{A\dot{B}}^{0}=\pi k_{A\dot{F}}\biggl(\overline{l}^{\dot{F}}\bullet\rho^{0}l^{E}+\sum_{n\neq0}2n^{2}\overline{a}^{\dot{F}}(n)\bullet\begin{pmatrix}0 & 0\\
1 & 0
\end{pmatrix}a^{E}(n)\biggl)k_{E\dot{B}},\\
\bigr(dx^{A\dot{B}}/d\overline{\tau}\bigl)_{C.M.}=\bigr(\overline{l}^{\dot{E}}\bullet l^{F}\,\overline{l}_{\dot{E}}\bullet l_{F}\bigl)^{-\frac{1}{2}}\overline{l}^{\dot{B}}\bullet l^{A},
\end{gather*}
to be co-directional as is the case for a point particle. The boundary
condition (\ref{eq:lrho1vanish}) is satisfied when $l_{1}^{A}$ is
set equal to $l_{2}^{A}$, in which case $\overline{l}^{\dot{F}}\bullet\rho^{0}l^{E}$
becomes equal to $\overline{l}^{\dot{F}}\bullet l^{E}$. According
to Equation (\ref{eq:l=00003Dk}), $k^{A\dot{B}}$ and $\overline{l}^{\dot{B}}\bullet l^{A}$
are co-directional, so the desired co-directionality is obtained when
\[
\overline{a}^{\dot{B}}(n)\bullet\begin{pmatrix}0 & 0\\
1 & 0
\end{pmatrix}a^{A}(n)
\]
is co-directional to $\overline{l}^{\dot{B}}\bullet l^{A}$. In this
case the active modes will increase the energy and mass of the string.

The point particle limit of the string action is obtained through
\begin{multline*}
c_{1}^{A}\thickapprox c_{2}^{A},\;d_{1\,A}^{*}\thickapprox d_{2\,A}^{*},\quad\partial_{1}c^{A}\thickapprox\partial_{1}d_{A}^{*}\thickapprox0,\quad h_{\alpha\beta}=diag(e(\tau)^{2},-1),
\end{multline*}
and when a kinetic term for $\phi$ is included in the string action
(\ref{eq:stringaction}), it becomes
\[
2\pi\int d\tau\bigl(\sqrt{lm}\,d_{A}^{\ast}\bullet\frac{dc^{A}}{d\tau}+c.c.-\frac{4l}{m}e(\tau)\,p^{\mu}p_{\mu}-lm\,e(\tau)^{-1}\bigl(\frac{d\phi}{d\tau}\bigr)^{2}\bigr).
\]
The equations of motion for $e(\tau)$ and $\phi(\tau)$ yield
\[
p^{\mu}p_{\mu}=m^{2}k^{2},\quad\frac{d\phi}{d\tau}=2k\,e(\tau),
\]
where $k$ is a constant of integration. Hence, the kinetic momentum
constraint results from  the joint action of the scalar field and
the metric without the need of specifying the mass of the particle. 

The octonionic generalization of the string action (\ref{eq:stringaction})
is
\begin{gather}
\int d\tau d\sigma\,e\bigr(\sqrt{lm}\:\overline{d}_{A}\times\rho^{\alpha}\nabla_{\alpha}c^{A}+o.c.-\frac{l}{m}det\,\bigl(d_{a}^{\dot{B}}\times\overline{d}_{a}^{A}\bigr)+lm\phi R(h_{\alpha\beta})\bigl),\label{eq:oct action}
\end{gather}
where we have written the worldsheet spinor index $a$ explicitly
because the matrix spinor notation conflicts with the correct order
of the octonion components. We assume that $c$ and $d$ are of the
form (\ref{eq:coordinatemomentumdefined}) so that the $\times$ products
in the action  (\ref{eq:oct action}) are octonionic. Since $\rho^{0}$
and $\rho^{1}$ are real, $\overline{d}_{A}$ transforms like $d_{A}^{\dagger}$
and $\rho^{\alpha}\nabla_{\alpha}c^{A}$ like $c^{A}$. The first
two terms in the Lagrangian therefore transform like the real part
of an octonionic spinor contraction and, according to Equation (\ref{eq:lorentz invariance contraction}),
are Lorentz invariant. From the expansion
\[
d_{a}^{\dot{B}}\times\overline{d}_{a}^{A}=d_{1}^{\dot{B}}\times d_{2}^{*A}+d_{2}^{\dot{B}}\times d_{1}^{*A}=(d_{1}^{\dot{B}}+d_{2}^{\dot{B}})\times(d_{1}^{*A}+d_{2}^{*A})-d_{1}^{\dot{B}}\times d_{1}^{*A}-d_{2}^{\dot{B}}\times d_{2}^{*A},
\]
it follows that the octonionic Hermitian matrix $d_{a}^{\dot{B}}\times\overline{d}_{a}^{A}$
transforms like $X$ in Equation (\ref{eq:XsXs}) with $S$ and $S^{\dagger}$
switched around, and its determinant is therefore Lorentz invariant.

In the octonionically generated ten-dimensional space-time, each of
the seven imaginary units generates a four-dimensional space-time.
Our universe appears to be confined to one of these, and therefore
we need a mechanism for dimensional reduction. It has been proposed
that the octonions could provide alternatives to compactification
of the extra dimensions \cite{key-14}. One possibility is to restrict
the endpoints of open strings to contain only one imaginary unit.
This would confine them to a four-dimensional space-time and would
limit the interaction between open strings with endpoints in different
four-dimensional space-times. The question is whether such a boundary
condition can be justified in other respects. Closed strings are not
subject to boundary conditions, and therefore all open strings are
able to interact gravitationally (assuming that closed strings mediate
gravity as in Superstring theory). If the ten-dimensional space-time
is homogeneous, then six out of seven strings that a string can interact
with gravitationally reside in the six other four-dimensional space-times.
This is an interesting ratio, since dark matter is thought to outweigh
visible matter roughly six to one.

\section{The quantum string}

Let $\Gamma$ be a space-like curve connecting two fixed points on
the boundaries of the worldsheet and let $\sigma^{\alpha}(u)$, $\sigma^{\alpha}(u')$
and $\sigma^{\alpha}(u'')$ be three points on this curve. We can
then define the worldsheet spinors
\begin{gather*}
\boldsymbol{d}_{\dot{B}}(u)\stackrel{def}{=}e\,v^{\alpha}\epsilon_{\beta\alpha}\rho^{\beta}d_{\dot{B}}(u),\quad\overline{\boldsymbol{d}}_{A}\mathrm{(u)}=e\,v^{\alpha}\epsilon_{\beta\alpha}\overline{d}_{A}(u)\rho^{\beta},\quad v^{\alpha}\stackrel{def}{=}\frac{d\sigma^{\alpha}}{du},
\end{gather*}
and the Clifford bracket
\begin{multline*}
\bigl\{ N^{'},M^{''}\bigr\}_{C.B.}\\
\stackrel{def}{=}\frac{1}{2\sqrt{\ell m}}\int_{\Gamma}du\,\bigl(\bigl\{\partial N^{'}/\partial c^{G},\partial M^{''}/\partial\overline{\boldsymbol{d}}_{G}\bigr\}+\bigl\{\partial N^{'}/\partial\overline{c}^{\dot{G}},\partial M^{''}/\partial\boldsymbol{d}_{\dot{G}}\bigr\}\\
-\bigl\{\partial M^{''}/\partial c^{G},\partial N^{'}/\partial\overline{\boldsymbol{d}}_{G}\bigr\}-\bigl\{\partial M^{''}/\partial\overline{c}^{\dot{G}},\partial N^{'}/\partial\boldsymbol{d}_{\dot{G}}\bigr\}\bigr)
\end{multline*}
for two functions $N$ and $M$ of $c^{A}$ and $\boldsymbol{d}_{\dot{B}}$
and their Dirac conjugates. $N$ and $M$ are worldsheet scalars;
so therefore their derivatives in the Clifford bracket are worldsheet
spinors and their Dirac conjugates. It is understood that these spinors
are to be contracted with each other so that the Clifford bracket
becomes a worldsheet scalar. Unprimed variables depend on $u$, and
variables with a single prime or a double prime depend on $u'$ and
$u''$ respectively.

The angular momentum current density defines the worldsheet scalars
\begin{gather*}
\mathrm{\boldsymbol{j}}_{AB}(u)\stackrel{def}{=}v^{\alpha}\epsilon_{\beta\alpha}\,\jmath_{AB}^{\beta}(u)=\sqrt{\ell m}\,\bigl(\overline{\boldsymbol{d}}_{A}(u)\bullet c_{B}(u)+\overline{\boldsymbol{d}}_{B}(u)\bullet c_{A}(u)\bigr),
\end{gather*}
with the Clifford brackets
\begin{gather*}
\bigl\{\mathrm{\boldsymbol{j}}_{AB}^{'},\mathrm{\boldsymbol{j}}_{EF}^{''}\bigr\}_{C.B.}=\bigl((\mathrm{\boldsymbol{j}}_{AE}^{'}\,\epsilon_{FB}+A\leftrightarrow B)+E\leftrightarrow F\bigr)\delta(u'-u''),\\
\bigl\{\mathrm{\boldsymbol{j}}_{AB}^{'},\mathrm{\boldsymbol{j}}_{\dot{E}\dot{F}}^{*''}\bigr\}_{C.B.}=0.
\end{gather*}
The quantization
\begin{gather*}
c^{A}\rightarrow\stackrel{>}{C^{A}},\quad d_{A}^{\ast}\rightarrow\stackrel{<}{D_{A}},\quad\mathrm{\boldsymbol{j}}_{AB}\rightarrow J_{AB},\\
\{N,M\}_{C.B.}\rightarrow\frac{1}{i\hslash}[N,M],
\end{gather*}
turns single spinors into infinite sets of spinors transforming like
ket and bra vectors under a unitary symmetry. $\mathrm{\boldsymbol{j}}_{AB}$
is hereby turned into an infinite-dimensional matrix $J_{AB}$ with
the commutation relations
\begin{gather}
[J_{AB}^{'},J_{EF}^{''}]=i\hslash\bigl((J_{AE}^{'}\,\epsilon_{FB}+A\leftrightarrow B)+E\leftrightarrow F\bigr)\delta(u'-u''),\quad[J_{AB}^{'},J_{\dot{E}\dot{F}}^{''\dagger}]=0.\label{eq:equal time JJ}
\end{gather}
Integrating both sides of Equation (\ref{eq:equal time JJ}) with
respect to both $u'$ and $u''$, we get the commutation relations
for the total Noether charges
\begin{gather}
[J_{AB}^{tot},J_{EF}^{tot}]=i\hslash\bigl((J_{AE}^{tot}\,\epsilon_{FB}+A\leftrightarrow B)+E\leftrightarrow F\bigr),\quad[J_{AB}^{tot},J_{\dot{E}\dot{F}}^{\dagger tot}]=0,\label{eq:com rel JJ total}\\
J_{AB}^{tot}\stackrel{def}{=}\int_{\Gamma}d\sigma^{\alpha}\epsilon_{\beta\alpha}J_{AB}^{\beta},\quad J_{AB}^{\beta}=\sqrt{lm}\,e\,(\overline{D}_{A}\bullet\rho^{\beta}C_{B}+\overline{D}_{B}\bullet\rho^{\beta}C_{A}),\nonumber 
\end{gather}
where $C_{A}$ and $\overline{D}_{A}$ are ket and bra vectors respectively.
Since $J_{AB}^{\alpha}$ is a conserved current density, its total
charge is independent of the path of integration.

In terms of
\begin{gather*}
N_{1}^{\dagger}\stackrel{def}{=}\frac{i}{4}(J_{22}^{tot}-J_{11}^{tot}),\quad N_{2}^{\dagger}\stackrel{def}{=}\frac{1}{4}(J_{11}^{tot}+J_{22}^{tot}),\quad N_{3}^{\dagger}\stackrel{def}{=}\frac{i}{2}J_{12}^{tot},
\end{gather*}
Equation (\ref{eq:com rel JJ total}) becomes the Lorentz algebra
\begin{gather*}
[N_{i},N_{j}]=i\hslash\epsilon_{ijk}N_{k},\quad[N_{i}^{\dagger},N_{j}^{\dagger}]=i\hslash\epsilon_{ijk}N_{k}^{\dagger},\quad[N_{i},N_{j}^{\dagger}]=0.
\end{gather*}
In tensor notation, the Lorentz generators can be written as
\[
J_{\mu\nu}=4\sigma_{\mu}^{A\dot{E}}\sigma_{\nu}^{B\dot{F}}(J_{AB}\epsilon_{\dot{E}\dot{F}}+J_{\dot{E}\dot{F}}^{\dagger}\epsilon_{AB}).
\]
The states of the string are therefore representations of the Lorentz
group, which is a minimum requirement for identifying them with elementary
particles.

It is well known that the space-time orbital angular momentum operator
has only integral eigenvalues. We shall show that the total angular
momentum of the spinor-space string also supports half-integral eigenvalues.
We consider only the term $n=1$ in the expansion  (\ref{eq:j0})
corresponding to the lowest mode:
\begin{multline}
\jmath_{AB}^{0,tot}\stackrel{def}{=}\int_{0}^{\pi}d\sigma\jmath_{AB}^{0}=-2i\pi\sqrt{lm}\,k_{A\dot{F}}\,\overline{a}^{\dot{F}}\bullet\begin{pmatrix}0 & 0\\
0 & 1
\end{pmatrix}a_{B}+A\leftrightarrow B\\
=-2i\pi\,\sqrt{lm}\,k_{A\dot{F}}\,a_{1}^{*\dot{F}}\bullet a_{2,B}+A\leftrightarrow B,\quad a_{B}=\Bigl(\begin{array}{c}
a_{1,B}\\
a_{2,B}
\end{array}\Bigr).\label{eq:j0tot}
\end{multline}
The quantized form of Equation (\ref{eq:j0tot}) is
\begin{equation}
J_{AB}^{0,tot}=-2i\pi\,\sqrt{lm}\,K_{A\dot{F}}\stackrel{>}{A_{2,B}}\bullet\stackrel{<}{A_{1}^{\dot{F}}}+A\leftrightarrow B,\label{eq:J0totquantum}
\end{equation}
where $J^{0}$ and $K$ are infinite-dimensional matrices, and $A_{1}$
and $A_{2}$ are infinite-dimensional vectors. Any matrix $M$ can
be decomposed according to
\begin{equation}
M=\stackrel{>}{V_{1}}\bullet\stackrel{<}{V_{1}}+i\stackrel{>}{V_{2}}\bullet\stackrel{<}{V_{2}}\label{eq:N decomposed}
\end{equation}
by factorizing its Hermitian and anti-Hermitian parts. When the Clifford
algebra is infinite-dimensional, the components of $V_{1}$ and $V_{2}$
can be chosen to be orthogonal to each other so that Equation (\ref{eq:N decomposed})
can be written as
\[
M=(\stackrel{>}{V_{1}}+i\stackrel{>}{V_{2}})\bullet(\stackrel{<}{V_{1}}+\stackrel{<}{V_{2}}).
\]
Consequently, any matrix can be written on the form $\stackrel{>}{A_{2}}\bullet\stackrel{<}{A_{1}}$.
When $K_{A\dot{F}}$ is invertible, it follows that Equation (\ref{eq:J0totquantum})
imposes no algebraic constraint on the total angular momentum of the
string. The only constraint is the Lorentz algebra which admits both
integral and half-integral eigenvalues. Note that this would not have
been the case if the spinor coordinates and momenta had been worldsheet
scalars instead of worldsheet spinors.

When $A_{2}=iA_{1}$, the total angular momentum can be expressed
in terms of two Hermitian matrices
\begin{equation}
J_{AB}^{0,tot}=2\pi\,\sqrt{lm}\,K_{A\dot{F}}A_{B}^{\;\;\;\dot{F}}+A\leftrightarrow B,\quad A_{B}^{\;\;\;\dot{F}}\stackrel{def}{=}\stackrel{>}{A_{1,B}}\bullet\stackrel{<}{A_{1}^{\dot{F}}},\label{eq:JKA}
\end{equation}
and we find that the Lorentz algebra (\ref{eq:com rel JJ total})
is solved by $K$ and $A$ being conjugate variables
\[
-2\pi\sqrt{lm}[A^{E\dot{F}},K_{A\dot{B}}]=i\hslash\delta_{A}^{E}\delta_{\dot{B}}^{\dot{F}},\quad[K_{A\dot{B}},K^{E\dot{F}}]=[A_{A\dot{B}},A^{E\dot{F}}]=0.
\]
This turns the angular momentum algebra into the space-time \textit{orbital}
angular momentum algebra which admits only integral eigenvalues.

When $A_{2}=A_{1}$, Equation (\ref{eq:JKA}) becomes
\[
J_{AB}^{0,tot}=-2\pi i\,\sqrt{lm}\,K_{A\dot{F}}A_{B}^{\;\;\;\dot{F}}+A\leftrightarrow B,
\]
and the Lorentz algebra (\ref{eq:com rel JJ total}) is solved by
the anticommutation relations
\[
-2\pi\sqrt{lm}\left\{ A^{E\dot{F}},K_{A\dot{B}}\right\} =\hslash\delta_{A}^{E}\delta_{\dot{B}}^{\dot{F}},\quad\left\{ K_{A\dot{B}},K^{E\dot{F}}\right\} =\left\{ A_{A\dot{B}},A^{E\dot{F}}\right\} =0.
\]
We expect this case to correspond to half-integral spin. When the
$K's$ anti-commute among themselves, the identity (\ref{eq:identity})
is no longer valid and the compatibility condition (\ref{eq:l=00003Dk})
for Equations (\ref{eq:c expansion}) and (\ref{eq:d expansion})
is replaced by the general condition
\begin{equation}
\frac{2\sqrt{lm}}{m^{2}}K_{A\dot{F}}L^{E\dot{F}}K_{E\dot{G}}K^{B\dot{G}}=\delta_{A}^{B}\cdot\underline{1},\quad L^{E\dot{F}}\stackrel{def}{=}\overline{L}^{\dot{F}}\bullet L^{E},\label{eq:KLKK}
\end{equation}
where $\overline{L}^{\dot{F}}$ and $L^{E}$ are ket and bra vectors
respectively. When the $K's$ commute among themselves, Equation (\ref{eq:KLKK})
reduces to the classical Equation (\ref{eq:l=00003Dk}).

\section{Conclusion}

We have examined the quantization of the point particle and string
in a spinorial space which forms a substructure of Minkowski space.
The classical point particle has been described by a unitarily invariant
system of integral curves in this space. This paves the way for quantization
in a more direct manner than the usual space-time based procedure.
We have obtained the Lorentz algebra for the quantum string from Poisson
brackets in spinor space and shown that  a spinor string can have
both integral and half-integral spin states. By enlarging the Clifford
algebra to an R-algebra tensor product with the octonions, we have
obtained a Lorentz invariant string action in ten-dimensional Minkowski
space. This lends support to the mathematical hypothesis that there
is a connection between the dimension of space-time and the normed
division algebras.

\section{Appendix}

To prove Equation (\ref{eq:lorentz invariance contraction}), we expand
$S$, $\chi$ and $\psi$ in terms of the octonion units:
\begin{gather*}
S_{\;\:A}^{E}=S_{\;\:A}^{E}(0)+S_{\;\:A}^{E}(k)e_{k},\\
\chi^{E}=\chi^{E}(0)+\sum_{i=1}^{7}\chi^{E}(i)e_{i},\quad\psi_{F}^{\ast}=\psi_{F}^{\ast}(0)+\sum_{j=1}^{7}\psi_{F}^{\ast}(j)e_{j},
\end{gather*}
where $S$ lies in the complex subspace corresponding to $e_{k}$.
When $det(S)$ is real, it follows from the general formula for determinants
\begin{gather*}
\epsilon_{AB}S_{\;\:E}^{A}S_{\;\:F}^{B}=det(S)\epsilon_{EF},\:\text{or}\:S_{\;\:E}^{A}S_{A}^{\;\:F}=-det(S)\delta_{E}^{F},
\end{gather*}
that
\begin{gather}
S_{\;\:E}^{A}(0)S_{A}^{\;\:F}(0)-S_{\;\:E}^{A}(k)S_{A}^{\;\:F}(k)=-det(S)\delta_{E}^{F},\nonumber \\
S_{\;\:E}^{A}(0)S_{A}^{\;\:F}(k)+S_{\;\:E}^{A}(k)S_{A}^{\;\:F}(0)=0.\label{eq:s determinant imaginary}
\end{gather}
Terms with imaginary units contribute to the expansion of the terms
in (\ref{eq:lorentz invariance contraction}) only if the product
of these units is $\pm1$. Terms with two or four imaginary units
can be computed $\mathit{as}$ $\mathit{if}$ $e_{n}e_{m}=-\delta_{nm}$
and $(e_{k}e_{i})(e_{j}e_{k})=\delta_{ij}$.

The terms on the l.h.s of Equation (\ref{eq:lorentz invariance contraction})
with no imaginary units are
\begin{gather}
-S_{A}^{\;\:F}(0)S_{\;\:E}^{A}(0)\psi_{F}^{\ast}(0)\chi^{E}(0)+o.c.\label{eq:no units}
\end{gather}
The terms with two imaginary units become
\begin{multline*}
\psi_{F}^{\ast}(0)S_{A}^{\;\:F}(0)S_{\;\:E}^{A}(k)\chi^{E}(k)+\psi_{F}^{\ast}(0)S_{A}^{\;\:F}(k)S_{\;\:E}^{A}(0)\chi^{E}(k)\\
+\psi_{F}^{\ast}(k)S_{A}^{\;\:F}(0)S_{\;\:E}^{A}(k)\chi^{E}(0)+\psi_{F}^{\ast}(k)S_{A}^{\;\:F}(k)S_{\;\:E}^{A}(0)\chi^{E}(0)\\
+S_{A}^{\;\:F}(0)S_{\;\:E}^{A}(0)\sum_{j=1}^{7}\psi_{F}^{\ast}(j)\chi^{E}(j)+\psi_{F}^{\ast}(0)S_{A}^{\;\:F}(k)S_{\;\:E}^{A}(k)\chi^{E}(0)+o.c.,
\end{multline*}
which, by use of Equation (\ref{eq:s determinant imaginary}) are
reduced to
\begin{gather}
S_{A}^{\;\:F}(0)S_{\;\:E}^{A}(0)\sum_{j=1}^{7}\psi_{F}^{\ast}(j)\chi^{E}(j)+S_{A}^{\;\:F}(k)S_{\;\:E}^{A}(k)\psi_{F}^{\ast}(0)\chi^{E}(0)+o.c.\label{eq:two imaginary units}
\end{gather}
The terms with three imaginary units become
\begin{multline}
-S_{A}^{\;\:F}(k)S_{\;\:E}^{A}(0)\sum_{i,j=1}^{7}\psi_{F}^{\ast}(j)\chi^{E}(i)(e_{j}e_{k})e_{i}\\
-S_{A}^{\;\:F}(0)S_{\;\:E}^{A}(k)\sum_{i,j=1}^{7}\psi_{F}^{\ast}(j)\chi^{E}(i)e_{j}(e_{k}e_{i})+o.c.,\label{eq:third power}
\end{multline}
(terms with two $e_{k}$'s vanish). $(e_{j}e_{k})e_{i}$ only contributes
if $e_{j}e_{k}=\pm e_{i}$, in which case we have $(e_{j}e_{k})e_{i}=e_{j}(e_{k}e_{i})$.
Together with Equation (\ref{eq:s determinant imaginary}), this makes
(\ref{eq:third power}) vanish. Terms with four imaginary units become
\begin{gather}
-S_{A}^{\;\:F}(k)S_{\;\:E}^{A}(k)\sum_{j=1}^{7}\psi_{F}^{\ast}(j)\chi^{E}(j)+o.c.\label{eq:four units}
\end{gather}
The total contribution to the expansion of the l.h.s. of Equation
(\ref{eq:lorentz invariance contraction}) is obtained by adding (\ref{eq:no units}),
(\ref{eq:two imaginary units}) and (\ref{eq:four units}), which
gives 
\begin{gather*}
det(S)\Bigl(\psi_{E}^{\ast}(0)\chi^{E}(0)-\sum_{j=1}^{7}\psi_{E}^{\ast}(j)\chi^{E}(j)\Bigr)+o.c.,
\end{gather*}
and is readily seen to be the same as the expansion of the r.h.s.
of Equation (\ref{eq:lorentz invariance contraction}).

\end{document}